\documentclass[aps,prl,twocolumn,superscriptaddress]{revtex4}
\usepackage{amsfonts}
\usepackage{amsmath}
\usepackage{amssymb}
\usepackage{graphicx}
\usepackage{epsfig}
\usepackage{bm}

\def\be{\begin{equation}}
\def\ee{\end{equation}}

\def\E{{\bf E}}

\def\F{{\bf F}}

\def\r{{\bf r}}

\providecommand{\U}[1]{\protect\rule{.1in}{.1in}}

\begin{document}
\preprint{ }
\title{Deterministic ratchet from stationary light fields}
\author{I. Zapata}
\affiliation{Dpto. F\'{\i}sica de Materiales, Universidad Complutense de Madrid,
E-28040 Madrid, Spain}
\author{S. Albaladejo}
\affiliation{Dpto. F\'{\i}sica de la Materia Condensada, Universidad Aut\'onoma de Madrid,
E-28049 Madrid, Spain}
\author{J. M. R. Parrondo}
\affiliation{Dpto. F\'{\i}sica At\'omica y Molecular, Universidad Complutense de Madrid,
E-28040 Madrid, Spain}
\author{J. J. S\'aenz}
\email{juanjo.saenz@uam.es}
\affiliation{Dpto. F\'{\i}sica de la Materia Condensada, Universidad Aut\'onoma de Madrid,
E-28049 Madrid, Spain}
\author{F. Sols}
\email{f.sols@fis.ucm.es}
\affiliation{Dpto. F\'{\i}sica de Materiales, Universidad Complutense de Madrid,
E-28040 Madrid, Spain}







\begin{abstract}
 Ratchets are dynamic systems where particle transport is induced by zero-average forces due to the interplay between nonlinearity and asymmetry. Generally, they rely on the effect of a strong external driving. We show that stationary optical lattices can be designed to generate particle flow in one direction while requiring neither noise nor driving. Such optical fields must be arranged to yield a combination of conservative (dipole) and nonconservative (radiation pressure) forces. Under strong friction all paths converge to a discrete set of limit periodic trajectories flowing in the same direction.
\end{abstract}
\volumeyear{2008}
\volumenumber{number}
\issuenumber{number}
\eid{identifier}
\startpage{1}
\endpage{ }
\maketitle

The combination of broken space inversion symmetry and strong external driving is becoming an important principle for the controlled motion of a variety of objects over a broad range of time and length scales.  The ratchet effect \cite{reimanRev,marc09}, whereby asymmetry and nonlinearity conspire to favor motion in one direction, has been
realized in an increasing number of systems such as  biological molecules \cite{vanOude99}, colloidal suspensions
\cite{prost,reiman_col},
liquid droplets \cite{linke},
superconducting vortices \cite{vicent} and devices \cite{zapa96,weis00}, cold atoms \cite{renzoni}, or magnetic domain
walls \cite{marconi}. A variety of questions has been explored theoretically, including the interplay between spatial asymmetries and thermal noise, time asymmetric forcing, topological driving, or
collective phenomena \cite{reimanRev,marc09}.

Despite the generality of these approaches, most previously
considered ratchets are based on some type of external driving,
random or periodic. The underlying reason is that spatial asymmetry is not
sufficient to induce directed motion: the necessary breaking of detailed balance requires also that the system be guided far from equilibrium. An external forcing is generally considered a necessary ingredient to produce the ratchet effect, which permits particle flow against friction in the absence of a net bias force.

There is however a way of driving a system out of equilibrium
which does not require external forcing and which has not been
considered in the ratchet literature, namely, the use of nonconservative stationary force fields such as those resulting from radiation pressure vortices \cite{cross}. Optical fields are easily tunable in general and affect any
polarizable object, from atoms to microscopic colloidal particles \cite{ashk97}. In this paper we show that a simple combination of null-average conservative and nonconservative steady forces can rectify the flow of damped particles. A deterministic ratchet stemming from purely stationary forces represents a novel concept in dynamics with considerable potential for fundamental and practical implications. For instance, it may permit the exploration of new forms of controlled atom motion in optical lattices \cite{cross,grynberg} and may be used to separate small particles with slightly different optical characteristics or friction properties \cite{Korda,Mac}.

The use of radiation pressure forces is reminiscent of the drift ratchet, which relies on the effect of a surrounding moving fluid \cite{reimanRev}. However, drift ratchets have been based so far on the use of some external driving. Nonconservative steady forces have been contemplated in the context of chemical motors but only in cases where the molecule spatial coordinate moves in a conservative field \cite{magn94,qian98}.


The two-dimensional field geometry arising in the
intersection region of two standing waves oriented along the $x$ and $y$ axes illustrates a particular simple case of nonconservative optical forces.
When the electric field is parallel to the $z$ axis, the interference between the standing waves pointing in perpendicular directions can lead to a lattice of radiation pressure vortices \cite{cross}. Specifically, for an electric field
\be
\E (\r) = E_0 \left[\sin (kx) + i \sin (ky) \right]\hat{\mathbf{z}} \, ,
\ee
the optical force on a polarizable particle becomes \cite{cross,alba09}
\begin{eqnarray}
\F_1 &=& \frac{\alpha' }{4}E_0^2
\left\{
\bm{\nabla} \left[\sin^2 (kx) +  \sin^2 (ky)\right] \right. \nonumber \\
&+&
\left.
 2\beta  \bm{\nabla} \times  \left[\cos (kx) \cos (ky)\hat{\mathbf{z}}\right]
\right\} \, ,
\label{force1-explicit}
\end{eqnarray}
where $\alpha(\omega)=\alpha '+i\alpha ''$ is the complex polarizability at frequency $\omega=ck$ ($c$ is the speed of light in the surrounding medium) and
$\beta \equiv \alpha''/\alpha'$. Notably, this force has a conservative (gradient) and a nonconservative (curl) contribution. The gradient term corresponds to the dipole force \cite{ashk86}, while the curl term accounts for radiation pressure \cite{cross,Ashkin}.

We consider the dynamics of a particle of mass $m$ surrounded by a viscous medium creating a friction coefficient $\eta$. In the overdamped limit ($\sqrt{m\alpha'}E_{0}k\ll\eta$) the equation of motion reads
\begin{equation}
\mathbf{\dot{r}}=\mathbf{F(r)}/\eta
\, , \label{eqnLangevin}%
\end{equation}
where $\mathbf{F(r)}$ is the stationary force field. We neglect here the fluctuating force due to thermal noise.
This is justified for low temperatures $T$ such that $k_{B}T\ll |\alpha| E_{0}^{2}$, which is the scale of the energy barriers constraining the particle motion.
The general effect of a nonzero temperature will be that of softening the otherwise sharp separation between the various flow regimes. Equation (3) allows us to treat the particle velocity at a given point as parallel to the local optical force.

\begin{figure}
[ptb]
\begin{center}
\includegraphics[width=6.4636in]
{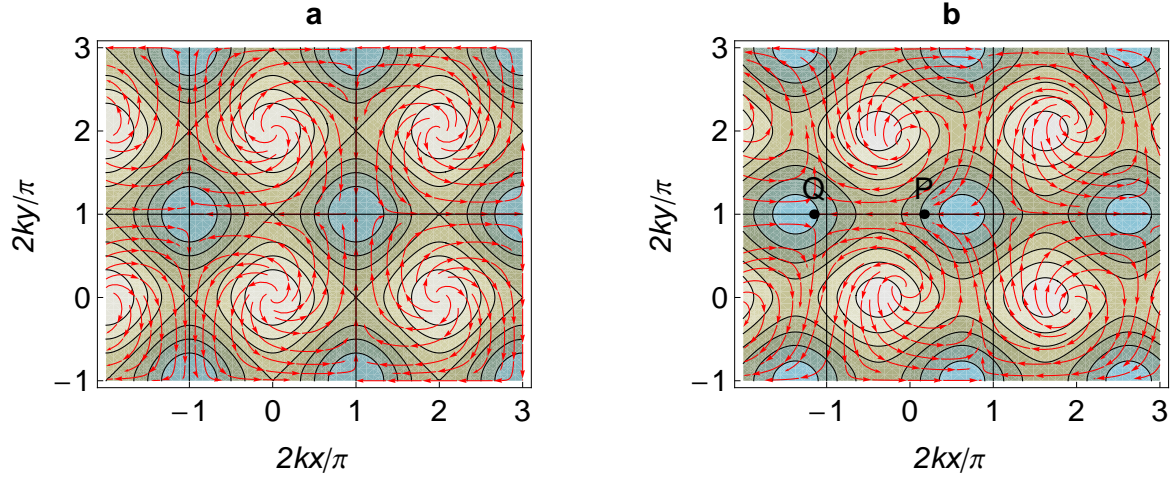}%
\caption{Equipotential lines are plotted with interstitial colors growing from blue (potential minima) to white (maxima). Red arrows show the individual trajectories of particles subject to the total force field, which includes the conservative force deriving from the shown potential and a nonconservative force obtained from the curl of a not shown vector potential.
The square $(-1,3)\times(-1,3)$ is a unit cell of the resulting periodic structure. Points P and Q are described in the caption of Fig. \ref{figZeroForceCurvesAndTrajectories}.
(a) Single-laser case [$\delta=0,\beta=1.4$ in Eqs. (\ref{force1-explicit}) and (\ref{force2-explicit})]; particles are confined within a quarter unit cell but even a minor amount of thermal noise allows them to diffuse isotropically throughout the periodic structure. (b) Double-laser ($\delta=1,\beta=1.4$); all trajectories evolve to the left, showing a ratchet effect which stems from purely deterministic and stationary forces.}
\label{figContourAndTrajectories}
\end{center}
\end{figure}

In Fig. \ref{figContourAndTrajectories}a we show the contour lines for the potential generating the gradient force (first term) in Eq. (\ref{force1-explicit}) and the velocity flow map under the total force $\F_1$ given in (\ref{force1-explicit}). The square $(-1,3)\times (-1,3)$ is the unit cell of the resulting periodic structure in the used reduced units.
The vortex centers (where flux lines start) coincide with the potential maxima, i.e. the nodes of the electric field. For $\beta<1$ pressure forces are too weak to free particles from their confinement within a quarter unit cell. In the case shown ($\beta=1.4$), particles are able to reach the boundaries of the quarter unit cells and move along the lines $2kx/\pi=2n+1$ and $2ky/\pi=2p+1$ with $n$ and $p$  integers.
Due to the absence of activation barriers for $\beta > 1$, a minor degree of thermal noise suffices to permit particles jumping between cells.
As a result, particles diffuse throughout the lattice without a privileged direction.

The long-time diffusion properties change qualitatively if a second laser is added within a suitable design. We consider the insertion of an independent laser which contributes a force
\be
\F_2 = \delta \, \frac{\alpha' }{4}E_0^2
\bm{\nabla} \left[\sin^2 (k'x+\phi)\right] \, .
\label{force2-explicit}
\ee
The dimensionless parameter $\delta$ gives the relative value between the intensity of the second laser and that of the conservative contributions of the first one [see Eq. (\ref{force1-explicit})]. Except for a final remark, we assume hereafter that the two lasers have the same wave number, $k'=k$.

A most interesting case is that of $\delta=1$ and phase shift $\phi=3\pi/8$, which breaks the $x\rightarrow -x$ inversion symmetry. Fig. \ref{figContourAndTrajectories}b reveals the dynamics under the total force field $\F=\F_1+\F_2=(F_x,F_y)$ for $\beta=1.4$ [see Eqs. (\ref{force1-explicit}) and (\ref{force2-explicit})]. The contour lines display the potential generating the gradient contribution to the total force $\F$, while the flow map shows the total force (or velocity) field, including the nonconservative forces [curl term in Eq. (\ref{force1-explicit})]. The vortex centers do not coincide with the potential maxima any more. Inspection of the flow lines reveals the remarkable fact that all trajectories are open and point asymptotically to the left: deterministic ratchet dynamics from stationary forces is possible.

\begin{figure}
[ptb]
\begin{center}
\includegraphics[
width=6.4636in
]%
{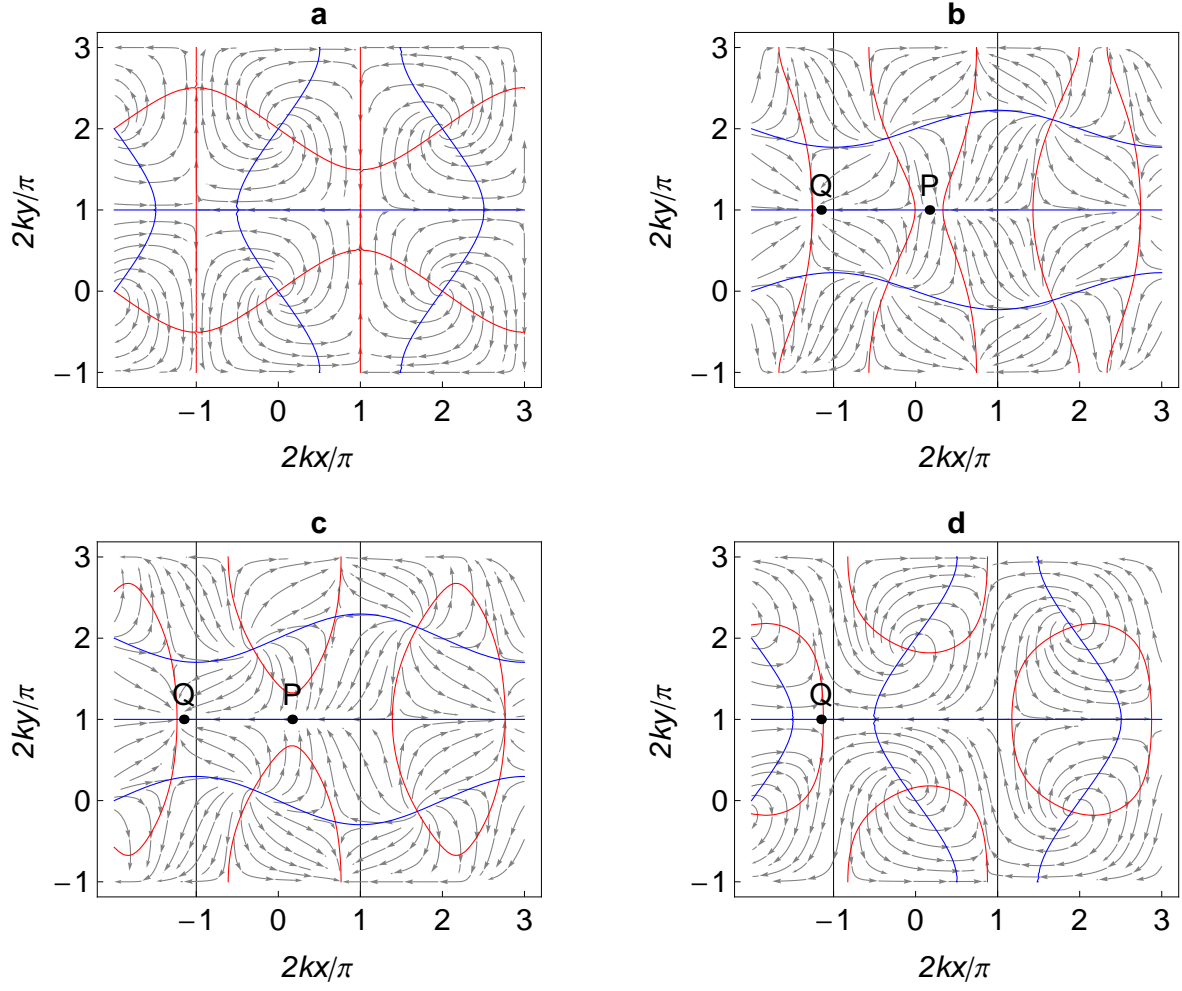}%
\caption{Grey arrows show particle trajectories under the effect of the total (nonconservative plus conservative) force.
The force is horizontal/vertical at the blue/red lines. (a) $\delta=0$, $\beta=1.4$; (b) $\delta=1$, $\beta=0.35$; (c) $\delta=1$, $\beta=0.45$; (d) $\delta=1$, $\beta=1.4$. Points P and Q become critical (zero force) at the transition between flow regimes, defined by the presence or absence of the various classes of attractive fixed points.
}%
\label{figZeroForceCurvesAndTrajectories}%
\end{center}
\end{figure}

Figure \ref{figZeroForceCurvesAndTrajectories} permits a more detailed analysis of the circulation patterns in the presence of the two lasers. The red/blue lines contain the points where the $x$/$y$ components of the total force vanish. Their intersections define the fixed points of the flow map, where $\F=0$. Figures \ref{figZeroForceCurvesAndTrajectories}a and \ref{figZeroForceCurvesAndTrajectories}d contain the same parameter values as Figs. \ref{figContourAndTrajectories}a and \ref{figContourAndTrajectories}b, while Figs. \ref{figZeroForceCurvesAndTrajectories}b and \ref{figZeroForceCurvesAndTrajectories}c display intermediate regimes corresponding to $\delta=1$ and $\beta$ values 0.35 and 0.45 respectively.

A feature which is common to all maps in Figs. \ref{figContourAndTrajectories} and \ref{figZeroForceCurvesAndTrajectories} is that the vertical component of the total force vanishes along the horizonal lines $2ky/\pi=2p+1$ with $p$  integer, which prevents vertical diffusion. We are thus entitled to focus on the action in the interval $2ky/\pi\in(-1,3)$.
On the contrary, the horizontal component vanishes at the vertical lines $2kx/\pi=2n+1$ ($n$ integer) only in the absence of the second laser ($\delta=0$, see Figs. \ref{figContourAndTrajectories}a and \ref{figZeroForceCurvesAndTrajectories}a). The most important effect of the second laser is that of rendering $F_x$ nonzero at the vertical lines separating the various quarter unit cells. This is clear in Figs. \ref{figContourAndTrajectories}b and \ref{figZeroForceCurvesAndTrajectories}b-d. Although the particle polarizability is not the easiest parameter to tune in a typical experiment, in the following we study the circulation trends as a function of $\beta$ because, against the background of a fixed conservative force, a clearer analysis is possible.

Figure \ref{figZeroForceCurvesAndTrajectories}b shows a case where nonconservative forces are too weak ($\beta=0.35$) to permit diffusion. Particles can move across squares but they always end up in an attractive equilibrium point. There are two types of them, located slightly to the right/left of representative points P/Q in the figure. An important change occurs for $\beta=0.393$, which is the root of the equation $F_x(x,\pi/2)=0$, $\partial_x F_x(x,\pi/2)=0$. At that value of $\beta$ the system undergoes a saddle-node bifurcation, as the attractive fixed points near P merge into point P and disappear; the short horizontal stretch comprising point P, with velocity pointing to the right, shrinks to zero. However, for values of $\beta$ only moderately above this first critical value (0.393), diffusion is still not possible. This is clearly shown in Fig. \ref{figZeroForceCurvesAndTrajectories}c ($\beta=0.45$), where all trajectories die in the attractive fixed points of the near-Q type, which still remain.

A major, qualitative change occurs at the critical value $\beta_c=1.03$, which is the root of the equation $F_x(x,\pi/2)=0$, $\partial_y F_y(x,\pi/2)=0$. At $\beta=\beta_c$
there is a
subcritical pitchfork bifurcation changing the
stability of the representative fixed point near Q, which is not attractive any more but rather becomes a saddle point (attractive in the $x$ direction and repulsive in the $y$ direction). A typical case is shown in Fig. \ref{figZeroForceCurvesAndTrajectories}d, where $\beta=1.4$. The result is that no attractive fixed points exist any longer. Particles can move throughout the lattice but, very importantly, only towards the left and in a deterministic manner (i.e. without requiring thermal activation). From stationary forces of null average we have designed a ratchet system which requires neither noise nor driving.

\begin{figure}
[ptb]
\begin{center}
\includegraphics[
width=6.4636in
]%
{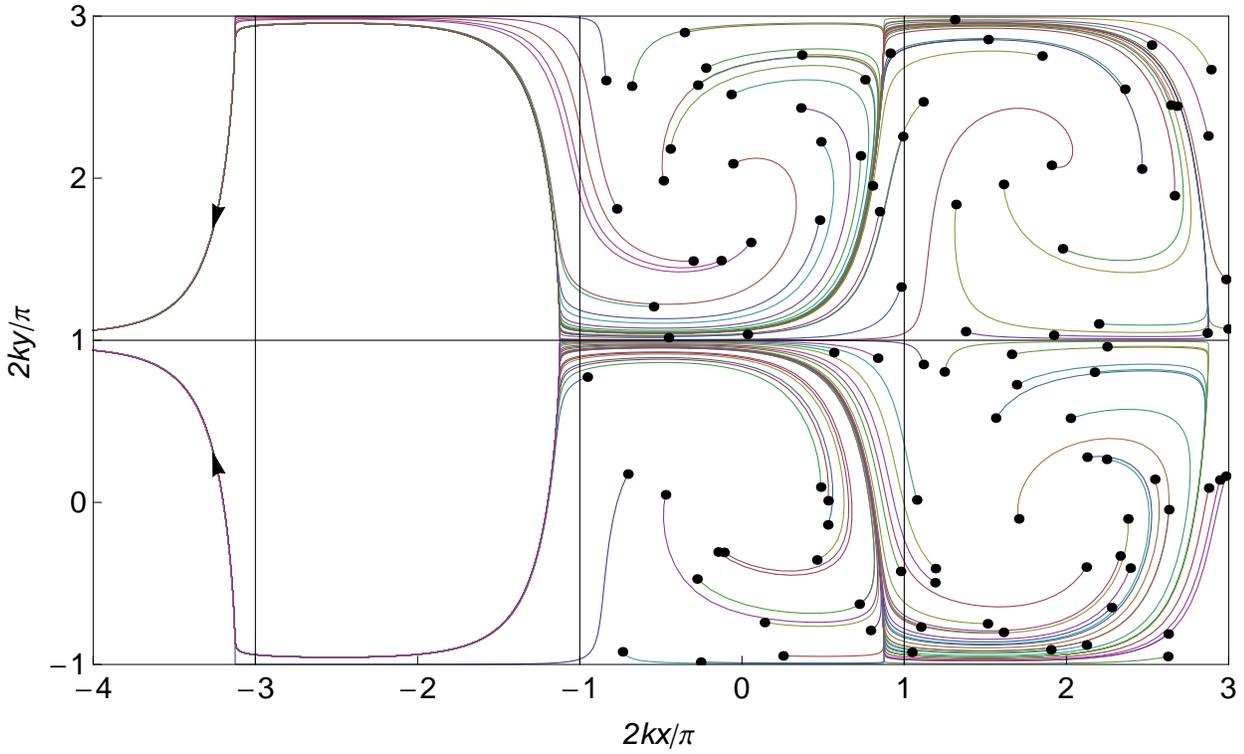}%
\caption{A set of 100 trajectories is shown for the setup of Figs. \ref{figContourAndTrajectories}b and \ref{figZeroForceCurvesAndTrajectories}d. Initial positions are random within the chosen unit cell. All trajectories converge to a limit periodic trajectory which flows to the left.}%
\label{figRandomAndAsymptoticTrajectories}%
\end{center}
\end{figure}

The deterministic character of the resulting ratchet is best appreciated in Fig. \ref{figRandomAndAsymptoticTrajectories}, where 100 trajectories are shown with initial positions chosen at random within the unit cell.
From a generalization to two-dimensional manifolds of the Poincar\'{e}-Bendixson theorem \cite{Hartman}, and as can be clearly seen in Figs. \ref{figZeroForceCurvesAndTrajectories}d and \ref{figRandomAndAsymptoticTrajectories}, limit attracting periodic trajectories exist.
These trajectories satisfy the differential equation $dy/dx=F_y/F_x$ with periodic boundary conditions.

\begin{figure}
[ptb]
\begin{center}
\includegraphics[
width=3.3797in
]%
{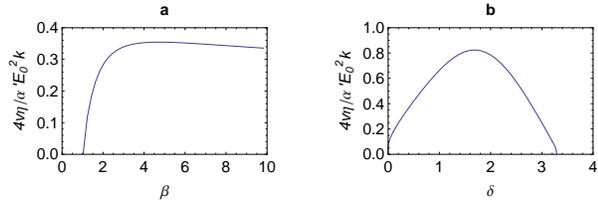}%
\caption{Average velocity $v$ of the limit periodic trajectories, in units of $\alpha'E_0^{2} k /4\eta$ [see Eqs. (\ref{force1-explicit})-(\ref{force2-explicit})]. (a) $v/(1+2\beta)$ as a function of $\beta$ for second-laser intensity $\delta=1$. (b) $v$ as a function of $\delta$ for $\beta=1.4$.
}%
\label{figAverageVelocity}%
\end{center}
\end{figure}

Figure \ref{figAverageVelocity}a shows the average velocity of the limit trajectories as a function of the parameter $\beta$, normalized to the amplitude of the total force (proportional to $2\beta+1$). The minimum value $\beta_c=1.03$ is required to have a nonzero mean velocity (ratchet effect). A detailed analysis shows that, for $\beta$ slightly above $\beta_c$, the average velocity grows linearly, $v(\beta)\sim \beta-\beta_c$. For large values of $\beta$, the velocity grows slower than linear, $v\simeq \pi \beta/2\ln \beta$.

Figure \ref{figAverageVelocity}b shows the average velocity for fixed $\beta=1.4$ as a function of the parameter $\delta$, which gives the relative strength of the second laser and is thus easier to tune. The limiting behaviors are $v\sim (\ln \delta)^{-1}$ for $\delta \rightarrow 0^{+}$, and $v\sim \sqrt{\delta -\delta_c}$ for $\delta \rightarrow \delta_c^{-}$, where $\delta_c=3.29$ gives the maximum intensity of the second laser permitting rectification of motion [see Eq. (\ref{force2-explicit})]. For $\delta > \delta_c$ conservative forces dominate and the ratchet effect disappears.

We have focused our study on the case $k'=k$ and $\delta=1$ in Eq. (\ref{force2-explicit}), which has the advantage of involving a single optical wavelength. Other conservative forces are of course possible. We note here that the case $k'=2k$, $\delta=1/4$, and $\phi=0$, which often stands as a paradigmatic ratchet potential \cite{reimanRev}, also yields a deterministic rectification of motion under the action of radiation pressure from the optical vortex lattice.

We conclude that suitably designed optical lattices can provide a playground for novel forms of nonconservative stationary dynamics. As a remarkable example, we have demonstrated theoretically that deterministic ratchets from stationary light forces are possible. Figure \ref{figRandomAndAsymptoticTrajectories} offers a vivid picture of how properly tailored steady fields can drag particles without a net bias force. The fine sensitivity of the flow regime to polarization and friction properties can pave the way to novel protocols of separation of different particle species \cite{Korda,Mac}.
We may generally expect the new dynamic concept here unveiled to open new vistas in the controlled transport of polarizable neutral bodies such as cold atoms or small dielectric particles.

We thank M. Marqu\'es for helpful discussions. This work was supported by Spain's Plan Nacional de F\'{\i}sica (MICINN) grants FIS2007-65723, MOSAICO, and   Nanolight, and by Comunidad de Madrid grant Microseres-CM.






\clearpage

\bigskip
\bigskip%

\bigskip

\bigskip%

\bigskip%

\end{document}